\documentclass[sigconf]{acmart}

\input{arxiv24-lightning-ir.sty}
\graphicspath{{./arxiv24-lightning-ir-figures}}

\begin{document}
\def\bstitle{\includegraphics[height=1.6ex]{lightning-ir-logo.ai} Lightning~IR: Straightforward Fine-tuning and Inference of Transformer-based Language Models for Information Retrieval \gdef\bstitle{Lightning~IR: Straightforward Fine-tuning and Inference of Transformer-based Language Models for Information Retrieval}}
\title[Lightning IR]{\bstitle}

\author{Ferdinand Schlatt}
\orcid{0000-0002-6032-909X}
\affiliation{
\institution{Friedrich-Schiller-Universit{\"a}t Jena}
\city{Jena}
\country{Germany}
}
\email{ferdinand.schlatt@uni-jena.de}

\author{Maik Fr{\"o}be}
\orcid{0000-0002-1003-981X}
\affiliation{
\institution{Friedrich-Schiller-Universit{\"a}t Jena}
\city{Jena}
\country{Germany}
}
\email{maik.froebe@uni-jena.de}

\author{Matthias Hagen}
\orcid{0000-0002-9733-2890}
\affiliation{
\institution{Friedrich-Schiller-Universit{\"a}t Jena}
\city{Jena}
\country{Germany}
}
\email{matthias.hagen@uni-jena.de}

\begin{abstract}
A wide range of transformer-based language models have been proposed for information retrieval tasks. However, including trans\-form\-er-based models in retrieval pipelines is often complex and requires substantial engineering effort. In this paper, we introduce Lightning~IR, an easy-to-use PyTorch Lightning-based framework for applying transformer-based language models in retrieval scenarios. Lightning~IR provides a modular and extensible architecture that supports all stages of a retrieval pipeline: from fine-tuning and indexing to searching and re-ranking. Designed to be scalable and reproducible, Lightning~IR is available as open-source: \url{https://github.com/webis-de/lightning-ir}.
\end{abstract}

\keywords{Software framework, Retrieval pipeline, Retrieval experiments}

\ccsdesc[500]{Information systems~Retrieval models and ranking}


\copyrightyear{2025}
\acmYear{2025}
\setcopyright{rightsretained}
\acmConference[WSDM '25]{Proceedings of the Eighteenth ACM International Conference on Web Search and Data Mining}{March 10--14, 2025}{Hannover, Germany}
\acmBooktitle{Proceedings of the Eighteenth ACM International Conference on Web Search and Data Mining (WSDM '25), March 10--14, 2025, Hannover, Germany}
\acmDOI{10.1145/3701551.3704118}
\acmISBN{979-8-4007-1329-3/25/03}


\epstopdfsetup{outdir=./}
\makeatletter
\gdef\@copyrightpermission{
  \begin{minipage}{0.3\columnwidth}
    \href{https://creativecommons.org/licenses/by/4.0/}{\includegraphics[width=0.90\textwidth]{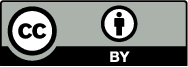}}
  \end{minipage}\hfill
  \begin{minipage}{0.7\columnwidth}
    \href{https://creativecommons.org/licenses/by/4.0/}{This work is licensed under a Creative Commons Attribution International 4.0 License.}
  \end{minipage}
  \vspace{5pt}
}
\makeatother

\maketitle

\section{Introduction}

Pre-trained transformer-based language models have become a cornerstone in information retrieval (IR) research~\cite{lin:2022}. Many different architectures have been proposed, each with their own implementation and training procedure. This plethora makes fine-tuning and comparing different model architectures cumbersome. However, all these models use similar backbones, are fine-tuned in the same way, and have only minor differences in their inference procedure.

To unify the usage of transformer-based language models in~IR, we present the \emph{Lightning~IR} framework. Lightning~IR builds on and extends PyTorch Lightning~\cite{falcon:2024} to provide several key features that set it apart from existing libraries for neural~IR: \Ni It is backbone agnostic, i.e., (almost) any HuggingFace~\cite{wolf:2020} transformer-based language model can be used. \Nii It supports the entire IR pipeline, from fine-tuning and indexing to searching and re-ranking. \Niii It is flexible and supports, for example, multi-vector or sparse bi-encoders and pointwise or listwise cross-encoders. \Niv It provides an easy to use~API and~CLI. \Nv It is highly configurable, allowing for reproducible experiments and painless model comparison. In this paper, we compare Lightning~IR to existing frameworks, describe its features and~API, and demonstrate Lightning~IR's capabilities.

\section{Comparison to Similar Frameworks}\label{sec:framework-comparison}

\begin{table}
  \centering
  \caption{Comparison of different IR~frameworks' supported stages (FT: fine-tuning, I: indexing, S: searching, RR: re-ranking) and model types (Bi-/Cr.-Enc.: bi-/cross-encoder, SV/MV: single-/multi-vector, DE: dense, SP: sparse, PW/LW: point-/listwise). (\cmark) denotes support for some model types.}
  \setlength{\tabcolsep}{2.2pt}
  \begin{tabular}{@{}lcccccccccc@{}}
    \toprule
    \bfseries Framework                     & \multicolumn{4}{c}{\bfseries Stages} & \multicolumn{6}{c}{\bfseries Model Types}                                                                                                                      \\
    \cmidrule(l@{\tabcolsep}r@{\tabcolsep}){2-5}\cmidrule(l@{\tabcolsep}){6-11}
                                            & FT                                   & I                                         & S        & RR     & \multicolumn{4}{c}{Bi-Enc.} & \multicolumn{2}{c}{Cr.-Enc.}                                     \\
    \cmidrule(l@{\tabcolsep}r@{\tabcolsep}){6-9}\cmidrule(l@{\tabcolsep}){10-11}
                                            &                                      &                                           &          &        & SV                          & MV                           & DE     & SP     & PW     & LW     \\
    \midrule
    baguetter~\cite{li:2024}                & \xmark                               & \cmark                                    & \cmark   & \xmark & \cmark                      & \xmark                       & \cmark & \xmark & \xmark & \xmark \\
    Capreolus~\cite{yates:2020}             & \xmark                               & \xmark                                    & \xmark   & \cmark & \cmark                      & \xmark                       & \cmark & \xmark & \xmark & \xmark \\
    Experimaestro-IR~\cite{piwowarski:2020} & \cmark                               & (\cmark)                                  & (\cmark) & \cmark & \cmark                      & \cmark                       & \cmark & \cmark & \cmark & \xmark \\
    FlexNeuART~\cite{boytsov:2020}          & \cmark                               & \xmark                                    & \cmark   & \cmark & \cmark                      & \xmark                       & \cmark & \xmark & \cmark & \xmark \\
    OpenNIR~\cite{macavaney:2020a}          & \cmark                               & \xmark                                    & \xmark   & \cmark & \cmark                      & \xmark                       & \cmark & \xmark & \cmark & \xmark \\
    Patapasco~\cite{costello:2022}          & \xmark                               & \cmark                                    & \cmark   & \cmark & \cmark                      & \xmark                       & \cmark & \xmark & \cmark & \xmark \\
    Pyserini~\cite{lin:2021a}               & \xmark                               & \cmark                                    & \cmark   & \cmark & \cmark                      & \xmark                       & \cmark & \xmark & \cmark & \xmark \\
    PyTerrier~\cite{macdonald:2021}         & \xmark                               & (\cmark)                                  & (\cmark) & \cmark & \cmark                      & \cmark                       & \cmark & \xmark & \cmark & \xmark \\
    RAGatouille\footnotemark[1]             & \cmark                               & \cmark                                    & \cmark   & \cmark & \xmark                      & \cmark                       & \cmark & \xmark & \xmark & \xmark \\
    rerankers~\cite{clavie:2024}            & \xmark                               & \xmark                                    & \xmark   & \cmark & \cmark                      & \cmark                       & \cmark & \cmark & \cmark & \cmark \\
    retriv\footnotemark[2]                  & \xmark                               & \cmark                                    & \cmark   & \xmark & \cmark                      & \xmark                       & \cmark & \xmark & \xmark & \xmark \\
    Seismic~\cite{bruch:2024}               & \xmark                               & \cmark                                    & \cmark   & \xmark & \cmark                      & \xmark                       & \xmark & \cmark & \xmark & \xmark \\
    SentenceBERT~\cite{reimers:2019}        & \cmark                               & \cmark                                    & \cmark   & \cmark & \cmark                      & \xmark                       & \cmark & \xmark & \cmark & \xmark \\
    Tevatron~\cite{gao:2022b}               & \cmark                               & \cmark                                    & \cmark   & \xmark & \cmark                      & \xmark                       & \cmark & \xmark & \xmark & \xmark \\
    \midrule
    Lightning~IR (Ours)                     & \cmark                               & \cmark                                    & \cmark   & \cmark & \cmark                      & \cmark                       & \cmark & \cmark & \cmark & \cmark \\
    \bottomrule
  \end{tabular}
  \label{tab:framework-comparison}
\end{table}

\footnotetext[1]{\url{https://github.com/AnswerDotAI/RAGatouille}}
\footnotetext[2]{\url{https://github.com/AmenRa/retriv}}

Several existing frameworks support fine-tuning and inference with neural retrieval models, but they differ in the supported stages of the retrieval pipeline and the types of models they can handle (see~Table~\ref{tab:framework-comparison} for an overview). Frameworks like RAGatouille, SentenceBERT~\cite{reimers:2019}, or Seismic~\cite{bruch:2024} focus on specific model types, while frameworks like PyTerrier~\cite{macdonald:2021} or Experimaestro-IR~\cite{piwowarski:2020} support several model types, but not all are available for all stages.

Instead, Lightning~IR is a framework that implements different models as generic, configurable, and extensible modules. In essence, a model is defined by its backbone encoder and how the contextualized embeddings of a query and document are combined to compute a relevance score. This design allows Lightning~IR to support a wide range of model types for all stages. In addition, as all model types use the same~API, they can use the same fine-tuning and inference procedures. This makes it easy to compare different model types and to experiment with new ones.

\section{Lightning~IR: Components}\label{sec:components}

Lightning~IR has four central components: \Ni model, \Nii dataset, \Niii trainer, and \Niv the CLI. We describe each component in detail in the following sections.

\subsection{Model}
\label{sec:model}

Lightning~IR supports two types of models: cross-encoders and bi-encoders. Both model types are built on top of backbone encoder models from HuggingFace~\cite{wolf:2020}. For example, the following code shows how to initialize a new bi-encoder and cross-encoder model using some pre-trained model from HuggingFace, where \texttt{\small{\{HF\_MODEL\}}} is a placeholder for the model name.

\begin{lstlisting}[language=PythonPlus,style=python]
from lightning_ir import BiEncoderModel, CrossEncoderModel
bi_encoder = BiEncoderModel.from_pretrained("{HF_MODEL}")
cross_encoder = CrossEncoderModel.from_pretrained("{HF_MODEL}")
\end{lstlisting}

A configuration class defines how each model type uses the contextualized embeddings generated by the backbone model to compute a relevance score. For example, a ColBERT-style model~\cite{khattab:2020} does not pool the contextualized embeddings. It instead uses late interaction to compute relevance scores over the contextualized query and document token embeddings. A ColBERT-style model can be initialized in Lightning~IR as follows.

\begin{lstlisting}[language=PythonPlus,style=python]
from lightning_ir import BiEncoderConfig
config = BiEncoderConfig(
  similarty_function="dot",
  query_pooling_strategy=None,
  doc_pooling_strategy=None,
  embedding_dim=128,
)
colbert = BiEncoderModel.from_pretrained(
  "bert-base-uncased", config=config
)
\end{lstlisting}

For usability and reproducibility, we combine a Lightning~IR model and tokenizer in a PyTorch Lightning module~\cite{falcon:2024}. The module is responsible for handling training and inference logic, but it also provides convenience functions to quickly score queries and documents. For example, the following code snippet shows how to compute scores using a pre-trained BERT-based bi-encoder and an ELECTRA-based cross-encoder.

\begin{lstlisting}[language=PythonPlus,style=python]
from lightning_ir import BiEncoderModule, CrossEncoderModule
bi_encoder = BiEncoderModule("webis/bert-bi-encoder")
cross_encoder = CrossEncoderModule("webis/monoelectra-base")
query = "What is the capital of Germany?"
docs = [
  "Berlin is the capital of Germany.",
  "Paris is the capital of France."
]
print(bi_encoder.score(query, docs).scores.numpy().round(2))
# [39.37 31.4]
print(cross_encoder.score(query, docs).scores.numpy().round(2))
# [ 7.81 -4.13]
\end{lstlisting}

\subsection{Dataset}
\label{sec:dataset}

Lightning~IR tightly integrates with \texttt{\small{ir\_datasets}}~\cite{macavaney:2021} to provide access to a wide range of common information retrieval datasets, but custom datasets are also supported. Datasets are split into four different classes: document, query, tuple, and run datasets. Document datasets iterate over a document collection and are used for indexing. Query datasets iterate over queries and are used for retrieval. Tuple datasets are used for fine-tuning as they contain samples consisting of a query and multiple documents. Run datasets contain ranked documents for a query (optional: relevance judgments) and are used for re-ranking; also fine-tuning is possible by sampling n-tuples from rankings. The following code snippet shows how to load the MS~MARCO passage dataset~\cite{bajaj:2018} and the TREC Deep Learning 2019 passage ranking dataset~\cite{craswell:2019}.

\begin{lstlisting}[language=PythonPlus,style=python]
from lightning_ir import DocDataset, QueryDataset, RunDataset
print(next(iter(DocDataset("msmarco-passage/train"))))
# DocSample(doc_id='0', doc='The presence of communication ...')
print(next(iter(QueryDataset("msmarco-passage/train"))))
# QuerySample(query_id='121352', query='define extreme')
print(RunDataset("msmarco-passage/trec-dl-2019", depth=3)[0])
# RankSample(query_id='1037798', query='who is robert gray', doc_ids=('7134595', '7134596', '8402859'), docs=(..., ..., ...))
\end{lstlisting}

Multiple datasets can be combined into a single PyTorch Lightning datamodule. Similar to a model's Lightning module, datamodules make fine-tuning and inference easier by handling data sampling and batching. They also cleanly separate fine-tuning and evaluation data. The following code snippet shows how to create a datamodule for fine-tuning a bi-encoder model on MS~MARCO triples and evaluating it on the TREC Deep Learning~2019 and~2020 passage ranking datasets.

\begin{lstlisting}[language=PythonPlus,style=python]
from lightning_ir import LightningIRDataModule
bi_encoder = BiEncoderModule(...)
data_module = LightningIRDataModule(
  module=module,
  train_dataset=TupleDataset("msmarco-passage/train/triples-v2"),
  inference_datasets=[
    RunDataset("msmarco-passage/trec-dl-2019"),
    RunDataset("msmarco-passage/trec-dl-2020"),
  ]
)
\end{lstlisting}

\subsection{Trainer}
\label{sec:trainer}

The trainer component builds on PyTorch Lightning's trainer class to provide flexible, scalable, and reproducible training. Lightning~IR adds functionality to support indexing, retrieval, and re-ranking. The following code snippet shows how to fine-tune a bi-encoder model on the MS~MARCO triples dataset. Hyperparameters (e.g., batch size, learning rate, number of epochs) should be adjusted in the module, datamodule, and trainer to the specific use case.

\begin{lstlisting}[language=PythonPlus,style=python]
from lightning_ir import LightningIRTrainer
module = BiEncoderModule("{HF_MODEL}", config=...)
data_module = LightningIRDataModule(train_dataset=...)
trainer = LightningIRTrainer(...)
trainer.fit(module, data_module)  
\end{lstlisting}

After fine-tuning a model, the trainer can be used to run inference for all stages of a retrieval pipeline. Indexing, searching, and re-ranking are all implemented as PyTorch Lightning callbacks (but indexing and searching are only needed for bi-encoders). Lightning~IR currently supports two indexing and searching methods: Faiss~\cite{douze:2024} for dense retrieval and a custom PyTorch-based~\cite{paszke:2019} sparse retrieval method. The following code snippet shows how to index and search documents using a fine-tuned bi-encoder model.

\begin{lstlisting}[language=PythonPlus,style=python]
from lightning_ir import FaissFlatIndexConfig, FaissFlatIndexer
module = BiEncoderModule("{PATH_TO_MODEL}")
data_module = LightningIRDataModule(
  inference_datasets=[DocDataset("msmarco-passage")]
)
index_callback = IndexCallback(
  index_dir="index", index_config=FaissFlatIndexConfig()
)
trainer = LightningIRTrainer(callbacks=[index_callback])
trainer.index(module, data_module)
\end{lstlisting}

To use an index for retrieval, the path of the index must be passed to a searcher class that matches the indexer class used to create the index. If a dataset has relevance judgments in \texttt{\small{ir\_datasets}} and evaluation metrics are specified in the module's configuration, the trainer will automatically evaluate the retrieval effectiveness. The following code snippet shows how to retrieve documents for a query using the indexed documents.

\begin{lstlisting}[language=PythonPlus,style=python]
from lightning_ir import FaissSearchConfig, SearchCallback
module = BiEncoderModule(
  "{PATH_TO_MODEL}", 
  evaluation_metrics=["nDCG@10"]
)
search_callback = SearchCallback("index", FaissSearchConfig(k=10))
data_module = LightningIRDataModule(inference_datasets=[QueryDataset("msmarco-passage/trec-dl-2019/judged")])
trainer = LightningIRTrainer(callbacks=[search_callback])
trainer.search(module, data_module)
\end{lstlisting}

\subsection{CLI}
\label{sec:cli}

Lightning~IR provides a command line interface~(CLI) to simplify the usage of the framework. The~CLI is built on top of PyTorch Lightning's~CLI and provides commands for fine-tuning, indexing, searching, and re-ranking. All options are configurable via command-line arguments or a configuration YAML~file. The configuration YAML~files are especially useful for reproducibility. For example, to fine-tune a bi-encoder model on the MS~MARCO triples dataset, the following command can be used.

\begin{lstlisting}[language=bash]
# > train.yaml
# trainer:
#   ... # trainer hyperparameters
# model:
#   class_path: BiEncoderModule
#   init_args:
#     model_name_or_path: bert-base-uncased
#     config:
#       class_path: BiEncoderConfig
#       init_args:
#         ... # model hyperparameters
# data:
#   class_path: LightningIRDataModule
#   init_args:
#     ... # data hyperparameters
#     train_dataset:
#       class_path: TupleDataset
#       init_args:
#         tuples_dataset: msmarco-passage/train/triples-v2
# optimizer:
#   class_path: torch.optim.AdamW
#   init_args:
#     ... # optimizer hyperparameters
lightning-ir fit --config train.yaml
\end{lstlisting}

The CLI also supports indexing, searching, and re-ranking via the \texttt{\small{index}}, \texttt{\small{search}}, and \texttt{\small{re\_rank}} commands. The configuration file and the command for indexing the MS~MARCO passage collection using a fine-tuned bi-encoder model are shown below. The configuration files for searching and re-ranking are similar.

\begin{lstlisting}[language=bash]
# > index.yaml
# trainer:
#   callbacks:
#   - class_path: IndexCallback
#     init_args:
#       index_dir: index
#       index_config:
#         class_path: FaissFlatIndexConfig
# model:
#   class_path: BiEncoderModule
#   init_args:
#     model_name_or_path: {PATH_TO_MODEL}
# data:
#   class_path: LightningIRDataModule
#   init_args:
#     inference_datasets:
#     - class_path: DocDataset
#       init_args:
#         doc_dataset: msmarco-passage
lightning-ir index --config index.yaml
\end{lstlisting}

\section{Lightning~IR: Supported Models}
\label{sec:supported-models}

Lightning~IR supports fine-tuning and running inference on a wide range of bi- and cross-encoder models but we have also added support for a number of popular models not natively fine-tuned in Lightning~IR. This includes all bi-encoders from the sentence transformers library~\cite{reimers:2019}, SPLADE models released by Naver labs~\cite{formal:2021a}, the official ColBERT checkpoints~\cite{khattab:2020}, and monoT5 and RankT5 models~\cite{nogueira:2020,zhuang:2022}. Further additional models can be easily added anytime by providing the corresponding configuration files and adding the model to the Lightning~IR model registry.

Table~\ref{tab:supported-models} compares a selection of supported models when re-ranking the TREC~2019 and~2020 Deep Learning track data~\cite{craswell:2019,craswell:2020} using the following configuration file and command.

\begin{lstlisting}[language=bash]
# > re-rank.yaml
# trainer:
#   logger: false
# model:
#   class_path: BiEncoderModule # or CrossEncoderModule
#   init_args:
#     model_name_or_path: {PATH_TO_MODEL}
#     evaluation_metrics:
#     - nDCG@10
# data:
#   class_path: LightningIRDataModule
#   init_args:
#     inference_datasets:
#     - class_path: RunDataset
#       init_args:
#         run_path_or_id: msmarco-passage/trec-dl-2019/judged
#     - class_path: RunDataset
#       init_args:
#         run_path_or_id: msmarco-passage/trec-dl-2020/judged
lightning-ir index --config re-rank.yaml    
\end{lstlisting}

\begin{table}
  \centering
  \setlength{\tabcolsep}{5pt}
  \caption{Effectiveness (nDCG@10) of a selection of models supported by Lightning~IR when re-ranking the TREC~2019 and~2020 Deep Learning track data.}
  \begin{tabular}{@{}lcccc@{}}
    \toprule
    \bfseries Model                       & \bfseries  TREC DL 2019 & \bfseries  TREC DL 2020 \\
    \midrule
    {\em Cross-Encoders}                                                                      \\
    \midrule
    monoELECTRA Large~\cite{schlatt:2024} & 0.750                   & 0.791                   \\
    monoT5 3B~\cite{nogueira:2020}        & 0.726                   & 0.752                   \\
    RankT5 3B~\cite{zhuang:2022}          & 0.721                   & 0.776                   \\
    \midrule
    {\em Bi-Encoders}                                                                         \\
    \midrule
    SBERT~\cite{reimers:2019}             & 0.705                   & 0.735                   \\
    ColBERT~\cite{formal:2021a}           & 0.732                   & 0.746                   \\
    SPLADE~\cite{khattab:2020}            & 0.715                   & 0.749                   \\
    \bottomrule
  \end{tabular}
  \label{tab:supported-models}
\end{table}

\section{Reproducibility Experiment}
\label{sec:reproducibility-experiment}

To demonstrate the capabilities of Lightning~IR, we conducted a reproducibility experiment. Using the \texttt{\small{bert-base-uncased}}%
\footnote{\url{https://huggingface.co/google-bert/bert-base-uncased}}
model as the backbone, we fine-tuned a single-vector bi-encoder~\cite{reimers:2019}, a SPLADE model~\cite{formal:2021a}, and a ColBERT model~\cite{khattab:2020}. The models are available in the HuggingFace model hub%
\footnote{\url{https://huggingface.co/webis/bert-bi-encoder}}\textsuperscript{,}%
\footnote{\url{https://huggingface.co/webis/splade}}\textsuperscript{,}%
\footnote{\url{https://huggingface.co/webis/colbert}}
along with the corresponding configuration files for reproducing the models and results using the Lightning~IR command-line interface.

\begin{table}
  \centering
  \setlength{\tabcolsep}{7pt}
  \caption{Effectiveness~(nDCG@10) of our fine-tuned models and of the official checkpoints for first-stage retrieval on TREC~DL~2019 and~2020. \sig~denotes a statistically significant difference ($p < 0.05$) between our and the original model.}
  \begin{tabular}{@{}lcccc@{}}
    \toprule
    \bfseries Model                        & \bfseries TREC DL 2019 & \bfseries  TREC DL 2020 \\
    \midrule
    SBERT (Ours)                           & 0.705                  & 0.696                   \\
    SBERT~\cite{reimers:2019} (Original)   & 0.705                  & 0.726                   \\
    \midrule
    SPLADE (Ours)                          & 0.760\kernSig          & 0.720\kernSig           \\
    SPLADE~\cite{formal:2021a} (Original)  & 0.722                  & 0.754                   \\
    \midrule
    ColBERT (Ours)                         & 0.738                  & 0.726                   \\
    ColBERT~\cite{khattab:2020} (Original) & 0.722                  & 0.723                   \\
    \bottomrule
  \end{tabular}
  \label{tab:reproducibility}
\end{table}

Table~\ref{tab:reproducibility} compares the effectiveness of our fine-tuned models with the official checkpoints%
\footnote{https://huggingface.co/sentence-transformers/msmarco-bert-base-dot-v5}\textsuperscript{,}%
\footnote{https://huggingface.co/naver/splade-v3}\textsuperscript{,}%
\footnote{https://huggingface.co/colbert-ir/colbertv2.0}
provided by the authors on the TREC~2019 and~2020 Deep Learning track data~\cite{craswell:2019,craswell:2020}. Our fine-tuned models achieve competitive effectiveness compared to the official checkpoints. Minor differences between our results and the official checkpoints can be attributed to randomness in the training process and are not statistically significant, except for the SPLADE models (our SPLADE model is more effective on TREC Deep Learning~2019 and less effective on TREC Deep Learning~2020). These results demonstrate that reproducing the effectiveness of state-of-the-art models is possible with minimal effort in Lightning~IR.

\section{Conclusion}\label{sec:sum}

We have introduced Lightning~IR, a PyTorch Lightning-based framework that enables straightforward fine-tuning and inference of transformer-based language models in retrieval tasks. By conducting a reproducibility experiment, we have demonstrated the capabilities of Lightning~IR and the simplicity and flexibility of its~API. With short code snippets and minimal effort, we were able to fine-tune and evaluate a variety of state-of-the-art models that achieve competitive effectiveness to their official checkpoints. Future work includes extending Lightning~IR to support additional efficient dense, sparse, and multi-vector indexing and retrieval pipelines, such as PLAID~\cite{santhanam:2022a} and Seismic~\cite{bruch:2024}, and to improve the latency and scalability of multi-vector and sparse retrieval models.

\bibliographystyle{ACM-Reference-Format-Abbreviated}
\bibliography{arxiv24-lightning-ir}

\end{document}